\documentclass{cimenics}
\usepackage[utf8]{inputenc}
\usepackage{amsfonts}
\usepackage{pstricks}
\usepackage{graphicx}
\usepackage{caption}
\usepackage{subcaption}

\title{CHARACTERIZING THE STRUCTURE OF COMPLEX \\
PROTEIN-PROTEIN INTERACTION NETWORKS}

\author{Allan A. Zea}

\heading{F. Author, S. Author y T. Author}

\address{\noindent \textit{allan.zea@grupo-quantum.org}\\
Escuela de Matemática, Facultad de Ciencias, UCV, Caracas, Venezuela

\author{Antonio Rueda-Toicen}

\address{\noindent\textit{antonio.rueda.toicen@algorithmicnaturelab.org}\\
Centro de Visualización de Imágenes, Instituto Nacional de Bioingeniería, UCV, Caracas, Venezuela
\\ Physics and Mathematics in Biomedicine Consortium, UCV, Caracas, Venezuela
\\ Algorithmic Nature Group, LABORES for the Natural and Digital Sciences, Paris, France}}

\abstract{ Network theorists have developed methods to characterize the complex interactions in natural phenomena. The structure of the network of interactions between proteins is important in the field of proteomics, and has been subject to intensive research in recent years, as scientists
have become increasingly capable and interested in describing the underlying structure of interactions in both normal and pathological biological processes. In this paper, we survey the graph-theoretic characterization of protein-protein interaction networks (PINs) in terms of structural features, and discuss its possible applications in biomedical research. We also perform a brief revision of network theory’s classical literature and discuss modern statistical and computational techniques to describe the structure of PINs. }

\keywords{Complex networks, biological network, protein-protein interaction, interactome, network science. }

\begin{document}

\section{INTRODUCTION}

Proteins are responsible of carrying out essential cellular processes like metabolism, vesicle transport, DNA transcription, among others. However, they rarely act alone: they must normally interact with other proteins in order to perform their functions. These physical interactions are important in several disciplines since they provide useful information regarding the function and development of certain diseases and health abnormalities \cite{Vidal,Barabasi1}.

\newpage

Many databases and public repositories have been recently launched in an attempt to finally assemble the whole set of interactions between proteins, for which it has become necessary to work on \textit{ad hoc} methodologies both to predict and model these interesting relationships.

In the following sections we survey modern techniques for characterizing the topology of protein-protein interaction networks and discuss the possible impact of such characterization in current biomedical research.

\section{UNDERSTANDING PROTEIN INTERACTIONS}

Protein-protein interactions (PPIs) are physical interactions between any two proteins that regulate the vast majority of biological processes within the cell. Predicting the whole set of PPIs for a given organism, therefore, comprises one of the major challenges in the field of proteomics, but the existing procedures for detecting and analyzing these interactions are still quite limited.

\subsection{Detection and analysis}

Systems biology has developed different approaches for identifying protein interactions \cite{Rao}. There are two main high-throughput methods to detect PPIs: namely, mass-spectrometry (MS) and methods based on yeast two-hybrid. The yeast two-hybrid (Y2H) approach allows \textit{in vivo} detection of protein interactions in yeast cells, and it is known to be more efficient and considerably less expensive than other techniques. This approach relies on the use of a transcription factor (originally Gal4, \cite{Fields}) that binds an upstream specific activating sequence (UAS) to activate a downstream reporter gene, which then leads to a specific phenotype, like growth on a selective medium or a color reaction \cite{Bruckner}. Nevertheless, although these biological assays have been implemented for various organisms in large-scale experiments \cite{CCSB}, their interactomes remain incomplete.

\begin{description} 

\item \textit{\textbf{Computational prediction of protein interactions.}} The detection of PPIs via high-throughput experimental techniques has slowed down in recent years in spite of the sharp increase in availability of genomic and proteomic data \cite{Singh1}, partly because of limitations with screening methods. For instance, Y2H experiments can produce false positives due to non-specific or ``promiscuous" interactions, and can undergo difficulties when trying to detect very transient ones \cite{Bruckner,Singh1}. Reasonable amounts of effort have thus been invested at predicting, curating and validating these uncertain interactions using computational techniques \cite{Zahiri}, some of which incorporate advanced insights from structural biology and machine learning (see, for example, \cite{Singh2}). A very popular tool for predicting PPIs is the \textit{Struct2Net} web server \cite{Singh1}, which is currently maintained by the computation and biology group at MIT's computer science laboratory and is freely available online.

\end{description}

\subsection{Protein interactions as graph-theoretic models}

A \textit{graph} is the pair $g:=(\mathcal{V},\mathcal{E})$ consisting of a vertex set $\mathcal{V}(g)$ and, correspondingly, an edge set $\mathcal{E}(g)\subseteq\{\{v_i,v_j\}\mid v_i,v_j\in\mathcal{V}(g),i\neq j\}$. Two vertices $v,w\in\mathcal{V}(g)$ are said to be \textit{adjacent} if there exists an edge $\{v,w\}\in\mathcal{E}(g)$ connecting them. In the context of PINs, two proteins are related if they establish a specific physical or biochemical interaction. Therefore, using this abstraction, we can represent the proteins in the interactome as vertices of a graph and the protein interactions (detected through Y2H or MS) as edges between them. These graphs or \textit{networks} can be easily constructed from the existing PPI annotations in most comprehensive repositories for interaction datasets such as the BioGRID \cite{Stark}. Figure 1 illustrates an example of protein interaction network. This network was built in Wolfram \textit{Mathematica} using the high-quality Y2H dataset of the \textit{CCSB Interactome Database} \cite{CCSB}.

\begin{figure}[ht]
\begin{center}
 \includegraphics[scale=0.75]{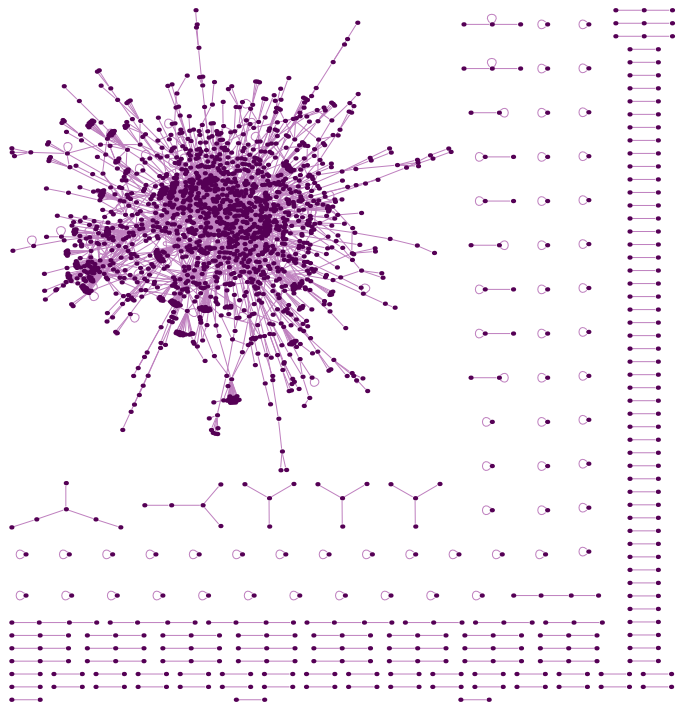}
 \caption{Network model for the PPI data of \textit{Saccharomyces cerevisiae}.}
 \label{figure1}
\end{center}
\end{figure}

\vspace{-4mm}

Some basic features regarding the structure of these PINs are shown in Table 1: The second and third columns respectively show the number of proteins in the network ($|\mathcal{V}(g)|$) and the total amount of detected interactions between them ($|\mathcal{E}(g)|$). The remaining columns, on the other hand, show the global and average clustering coefficients ($C$ and $\overline{C}$, respectively), which relate to the number of interactions between the neighbors of each protei, and the naive scaling exponent ($\gamma_{naive}$) that we will briefly describe in the next section.

\begin{table}[h]
\begin{center}\label{table1}
\caption{Some global features of the PINs in \cite{CCSB}.}
\begin{tabular}{cccccc}
  \hline
Protein interaction dataset & $|\mathcal{V}(g)|$ & $|\mathcal{E}(g)|$ & $\gamma_{naive}$ & $C$ & $\overline{C}$ \\
  \hline
\textit{Saccharomyces cerevisiae} & 2018 & 2930 & 1.79 & 0.02 & 0.04 \\
\textit{Caenorhabditis elegans} & 1493 & 1817 & 1.66 & 0.01 & 0.01 \\
\textit{Arabidopsis thaliana} & 4866 & 11374 & 1.69 & 0.03 & 0.09 \\
\textit{Homo sapiens} & 4303 & 13944 & 1.50 & 0.03 & 0.05 \\
  \hline
\end{tabular}
\end{center}
\end{table}

A compendium of other important measurements for describing the network's dynamics can be consulted in \cite{Albert,Newman1,Costa}.

\section{THE STRUCTURE OF PROTEIN INTERACTION NETWORKS}

\subsection{Degree Distribution}

An intriguing feature of a network's structure is its degree distribution. Many classical works have claimed that PINs are scale-free, which is to say that their degree distribution approximately follows a power-law behavior \cite{Albert,Barabasi2}. Thus, given some exponent $\gamma$, the probability $P(k)$ that a randomly chosen vertex in the PIN will have $k$ edges satisfies the asymptotic relation
\begin{equation}\label{eq1}
    P(k)\sim k^{-\gamma}
\end{equation}

\noindent that causes the degrees in the network to be distributed as shown in Fig. 2a. It is clear to observe from this figure that proteins with few interactions are significantly more frequent than those related with a large amount of other proteins. This characterization, to some extent, reflects the overly important role only a small number of proteins play in the cell's functionality, which is essential, for example, to understand the network's tolerance to failure and its vulnerability to targetted attack \cite{Iyer}.

\begin{figure}[ht]
\begin{center}
    \begin{subfigure}{.5\textwidth}
        \begin{center}
            \includegraphics[scale=0.9]{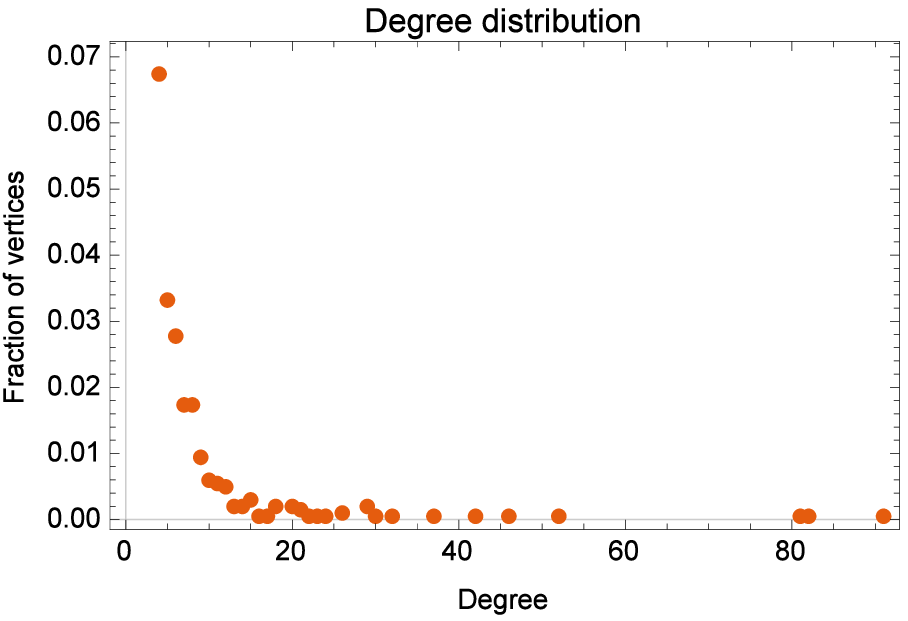}
            \caption{Corresponding degree distribution}
            \label{fig2:sub1}
        \end{center}
    \end{subfigure}%
    \begin{subfigure}{.5\textwidth}
        \begin{center}
            \includegraphics[scale=0.9]{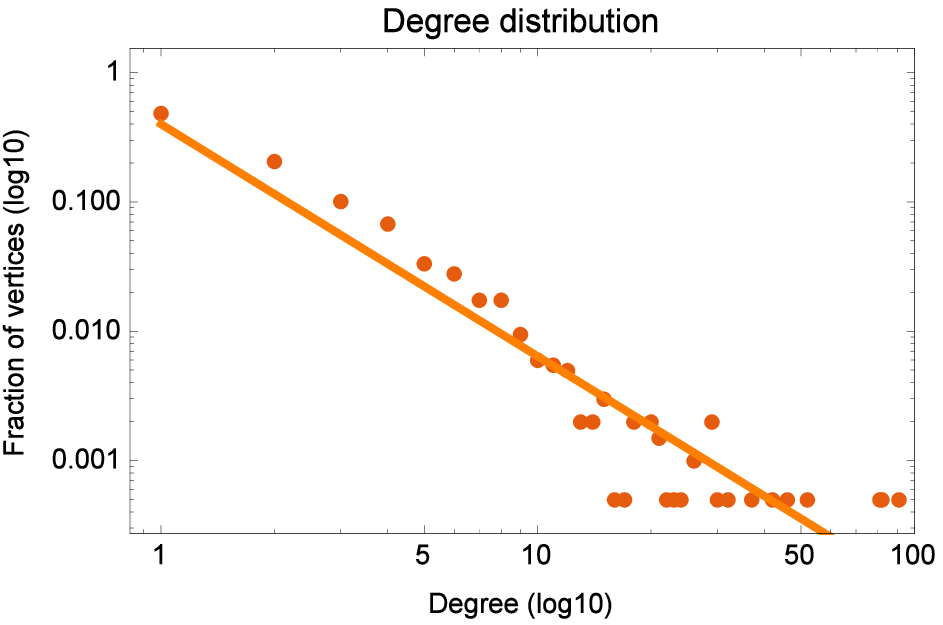}
            \caption{Fitted distribution in logarithmic scales}
            \label{fig2:sub2}
        \end{center}
    \end{subfigure}%
 \caption{Protein interaction network of \textit{Saccharomyces cerevisiae} \cite{CCSB}.}
 \label{figure2}
\end{center}
\end{figure}

\vspace{-4mm}

\begin{description}
\item \textit{\textbf{Estimating the scaling parameter.}} The scaling exponent in Eq. (\ref{eq1}) is often used as a measure for quantifying complexity in many real-world networks, that is normally calculated by fitting a line over the degree distribution plotted in logarithmic scales (Fig. 2b), where the slope is, presumably, $\gamma$. However, this procedure is known to be misleading, since a straight line in a log-log plot is not a sufficient condition for power-law behavior \cite{Clauset}. An efficient way to estimate this parameter is through the maximum-likelihood estimator (MLE) described in \cite{Clauset,Newman2}. In general, for discrete data distributions, this MLE is given by
\begin{equation}\label{eq2}
    \hat{\gamma}\simeq 1+{\left[\sum_{i=1}^n{\ln\frac{k_i}{k_{min}-\frac{1}{2}}}\right]}^{-1}
\end{equation}

\noindent where $k_{min}$ is the lower bound for the power-law behavior and $\hat{\gamma}\rightarrow\gamma$ in the limit of large $n$. 

\item \textit{\textbf{Testing the power-law hypothesis.}} The work of Stumpf and Ingram \cite{Stumpf}, in contrast to the claims of previous studies, argued that power-law distributions could not be fitted sufficiently well in much of the available PIN data. Furthermore, they found that the stretched exponential and log-normal distributions are often best fits for this empirical data. For this reason, Clauset et al. \cite{Clauset} also accompanied their approach with  statistics to test whether the power-law hypothesis for the empirical distribution under study is a plausible one.
\end{description}

\subsection{Centrality measures}

Let $g:=(\mathcal{V},\mathcal{E})$ be an undirected graph with a vertex set $\mathcal{V}(g)$, edge set $\mathcal{E}(g)$ and let $n=|\mathcal{V}(g)|$. We can represent graph $g$ by an $n\times n$ symmetric matrix $A=(a_{ij})$ called \textit{adjacency matrix}, where
\begin{equation}\label{eq3}
    a_{ij}=\left\{
        \begin{array}{ll}
            1 & \quad \textnormal{if vertices }i\textnormal{ and }j\textnormal{ are adjacent,} \\
            0 & \quad \textnormal{otherwise.}
        \end{array}
    \right.
\end{equation}

Given two vertices $v,w\in\mathcal{V}(g)=\{1,2,...,n\}$, we define the following centrality measures:

\begin{description}
\item \textit{\textbf{Degree centrality.}} The degree centrality $\mathcal{C}_{deg}$ of a vertex $v$ is the total number of vertices to which $v$ is connected. Formally, we have a mapping $\mathcal{C}_{deg}:\mathcal{V}(g)\rightarrow\mathbb{N}$ such that
\begin{equation}\label{eq4}
    \mathcal{C}_{deg}(v)=\sum_{j=1}^{n}{a_{vj}},
\end{equation}
where $\mathbb{N}$ is the set of natural numbers and $1\leq v\leq n$. Broadly speaking, a vertex $v$ is ranked as important by $\mathcal{C}_{deg}$ if it is connected to many other vertices. This is one of the most intuitive but still useful ideas of centrality, e.g. the most central vertex in a friendship network would be the person having the greatest amount of friends.

\item \textit{\textbf{Eigenvector centrality.}} Eigenvector centrality, $\mathcal{C}_{eig}$, is often referred as a more sophisticated formulation of degree centrality. Unlike degree centrality, a vertex $v$ is ranked as important by $\mathcal{C}_{eig}$ if it is connected to other vertices which are themselves important. Thus, we have:
\begin{equation}\label{eq5}
    \mathcal{C}_{eig}(v)=\frac{1}{\lambda}\sum_{w\in N(v)}{\mathcal{C}_{eig}(w)}=\frac{1}{\lambda}\sum_{v\in\mathcal{V}(g)}{a_{vw}\mathcal{C}_{eig}(w)},
\end{equation}
where $\lambda$ is a constant and $\mathcal{N}(v)=\{w\in\mathcal{V}(g)\mid a_{vw}=1\}$ is the set of neighbors of vertex $v$. Letting $\textnormal{\textbf{C}}=(\mathcal{C}_{eig}(1),\mathcal{C}_{eig}(2),...,\mathcal{C}_{eig}(n))$ denote the vector of centralities, we can write the above equation in matrix form as
\begin{equation}\label{eq6}
    \lambda\textnormal{\textbf{C}}=\textnormal{\textbf{A}}\textnormal{\textbf{C}},
\end{equation}
and therefore we see that $\textnormal{\textbf{C}}$ is an eigenvector of the adjacency matrix with eigenvalue $\lambda$. Eigenvector centrality is frequently used by search engines like Google to rank websites by relevance, because, as a matter of fact, websites are more likely to be visited if they are linked to other important websites that users on the Internet can reach.

\item \textit{\textbf{Betweenness centrality.}} Another important measurement of centrality in complex networks is betweenness, which is based upon the graph-theoretic notion of path. A path is a sequence of distinct vertices that pass over following edges accross a graph, from a vertex $v$ to some vertex $w$. Given $u\in\mathcal{V}(g)$ we define its betweenness centrality, $\mathcal{C}_{bet}(u)$, as follows:
\begin{equation}
    \mathcal{C}_{bet}(u)=\sum_{v\neq u\neq w\in\mathcal{V}(g)}{\frac{\sigma(v,u,w)}{\sigma(v,w)}},
\end{equation}
where $\sigma(v,w)$ is the total number of shortest paths between vertices $v$ and $w$, and $\sigma(v,u,w)$ is the total number of these paths passing through vertex $u$. Vertices with high betweenness centrality are often called bottlenecks. Betweenness is, in broad terms, a measure of the control these vertices conduct over the flow of information within the network.
\end{description}

Identifying and evaluating core proteins for an organism's interaction network has been one of the major goals of systems biology for several years. As a consequence, plenty of works have focused in the systematic study of centrality in PINs. Yu et al. \cite{Yu} found that there was a strong correlation between the ``bottleneckedness" of vertices in protein interaction networks and protein essentiality, i.e. bottlenecks are more likely to be essential: a fundamental observation for understanding lethality \cite{Jeong} and disease associations in PINs \cite{Vidal}.

\subsection{Attack robustness of complex networks}

Robustness is a fundamental issue in the study of complex networks, which concerns to the changes a networked system’s structure undergoes after a portion of its vertices are removed. These changes largely depend on the way degrees are distributed in the system as well as the vertex removal method, and can provide insight to study the network's capacity to withstand failure after some of its most essential components have been compromised or damaged.

\begin{description}
\item \textit{\textbf{Resilience to centrality attack.}} A recent paper by Iyer et al. \cite{Iyer} explored quantitative properties of robustness in complex network topologies and discussed their relationship to the notion of centrality. In particular, they focused in both random and centrality-based vertex removal strategies for degrading various empirical networks with different degree distributions. Their results suggested that, although networks with scale-free topology are tolerant to error (random elimination of poorly connected vertices), the effect of removing vertices with high centrality (degree, betweenness and eigenvector) is detrimental for the overall network's dynamics. However, they also pointed out that the degree distribution as well as the \textit{assortative mixing}, $r$, are determining parameters in a network's ability to resist these targeted attacks.

\item \textit{\textbf{Random networks: a critical comparison.}} Random networks have received a great deal of attention in network science. They are a special kind of network, whose vertices are more uniformly distributed than in scale-free networks (they resemble a Poisson distribution). Figure 3 shows sequential targeted attacks performed on a random network made of 756 vertices and 1685 edges. For this, we calculated the degree centralities for each vertex and, subsequently, removed a percentage of those with the highest $\mathcal{C}_{deg}$.
\end{description}

\begin{figure}[ht!]
\begin{center}
    \begin{subfigure}{.5\textwidth}
        \begin{center}
            \includegraphics[scale=0.6]{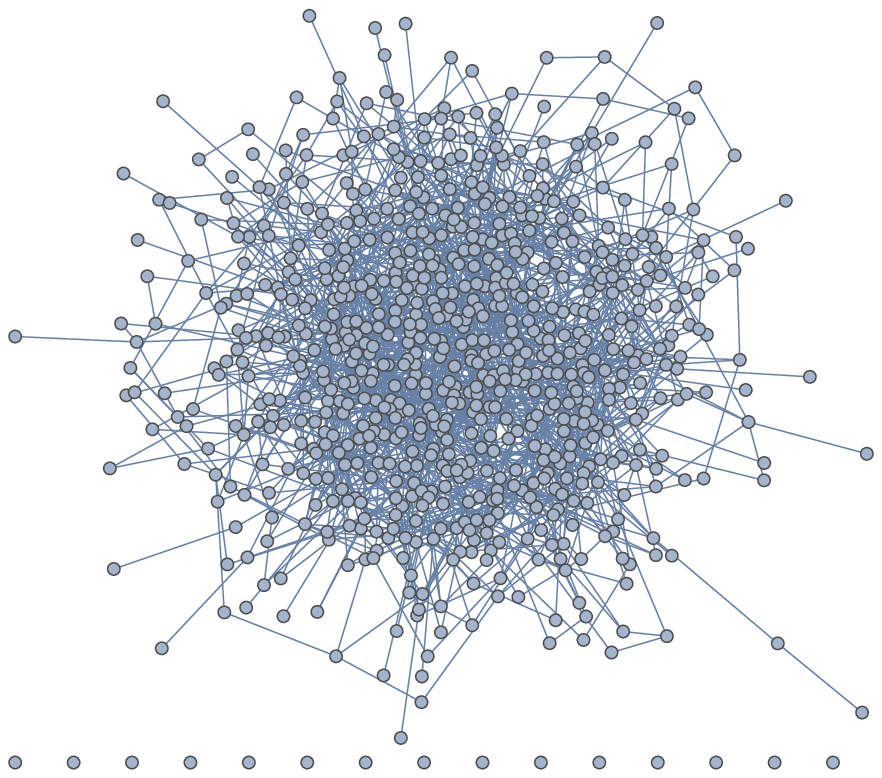}
            \caption{No vertices removed}
            \label{fig3:sub1}
        \end{center}
    \end{subfigure}%
    \begin{subfigure}{.5\textwidth}
        \begin{center}
            \includegraphics[scale=0.8]{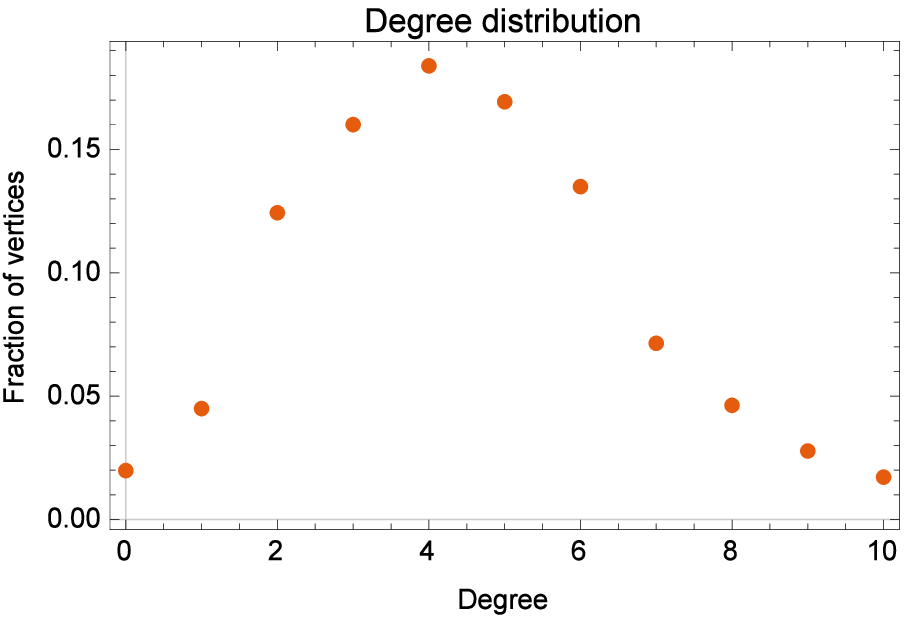}
            \caption{Corresponding degree distribution}
            \label{fig3:sub2}
        \end{center}
    \end{subfigure}
    \\
    \begin{subfigure}{.5\textwidth}
        \begin{center}
            \includegraphics[scale=0.65]{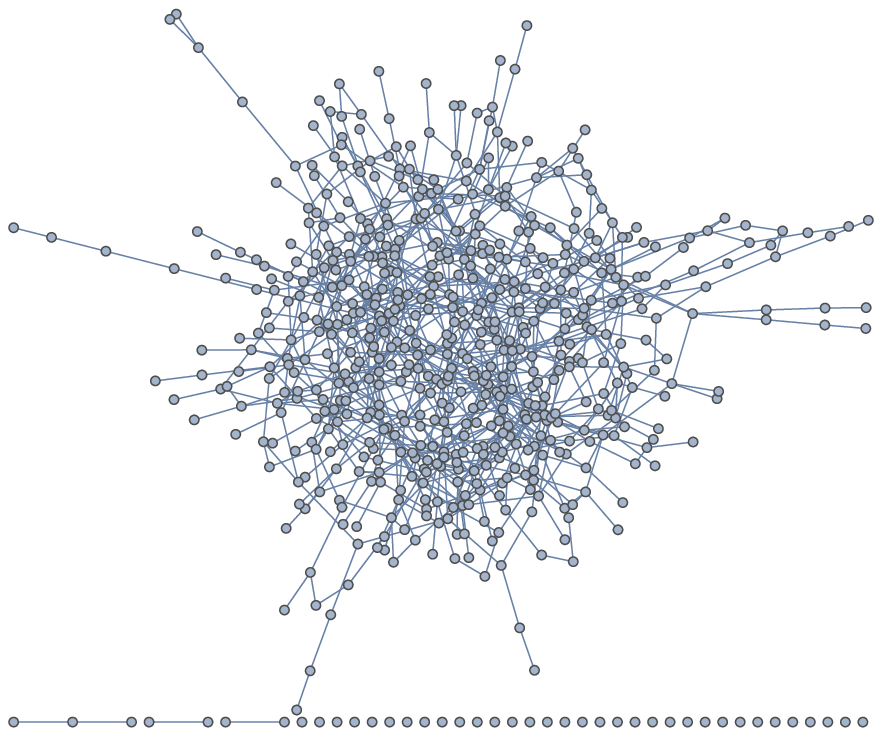}
            \caption{15\% of vertices removed}
            \label{fig3:sub3}
        \end{center}
    \end{subfigure}%
    \begin{subfigure}{.5\textwidth}
        \begin{center}
            \includegraphics[scale=0.8]{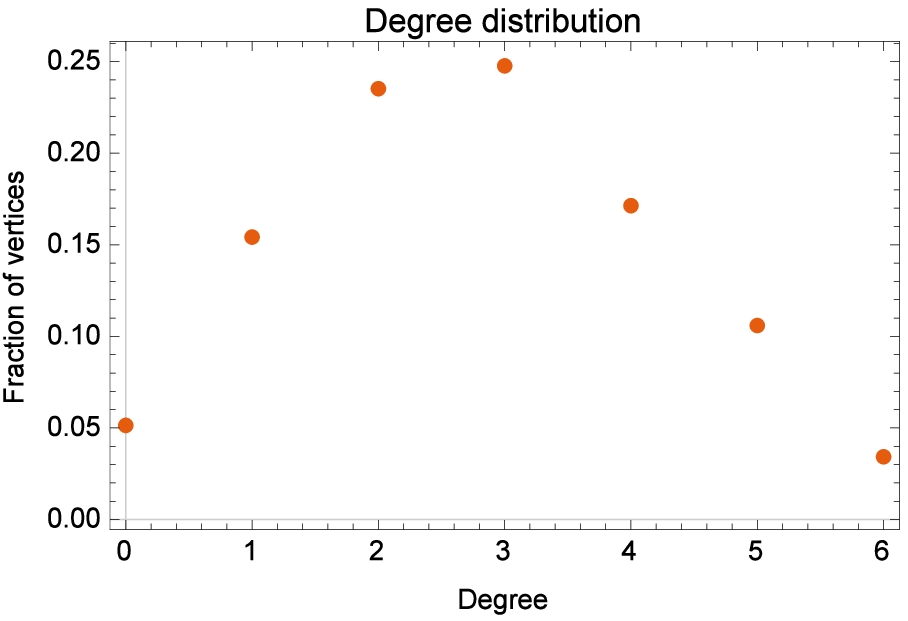}
            \caption{Modified degree distribution}
            \label{fig3:sub4}
        \end{center}
    \end{subfigure}%
    \\
    \begin{subfigure}{.5\textwidth}
        \begin{center}
            \includegraphics[scale=0.55,angle=270,origin=c]{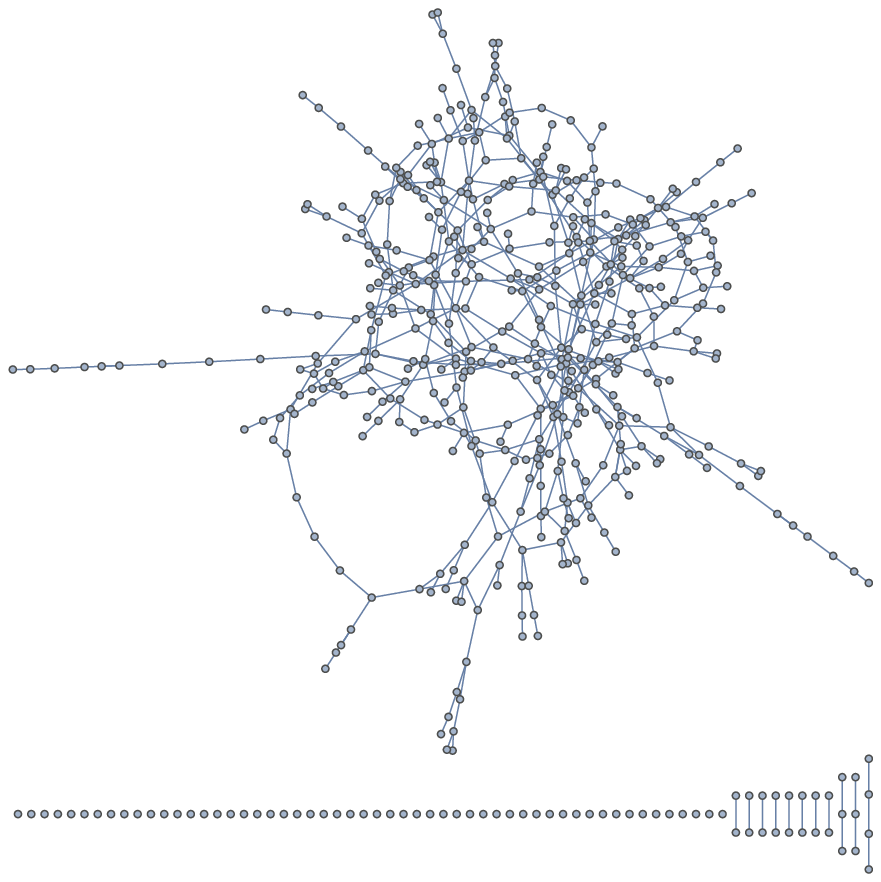}
            \caption{25\% of vertices removed}
            \label{fig3:sub5}
        \end{center}
    \end{subfigure}%
    \begin{subfigure}{.5\textwidth}
        \begin{center}
            \includegraphics[scale=0.8]{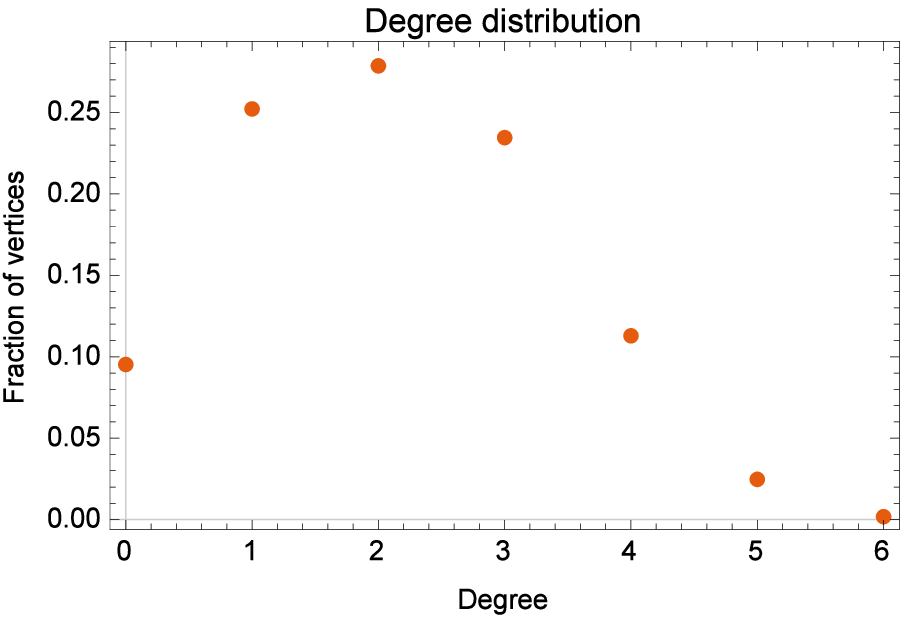}
            \caption{Modified degree distribution}
            \label{fig3:sub6}
        \end{center}
    \end{subfigure}%
 \caption{Sequential targeted attack to a random network. The left-hand side figures display changes in the network's structure after successful elimination of vertices with the highest degree centralities (hubs), while figures in the right show modifications in their degree distributions.}
 \label{figure3}
\end{center}
\end{figure}

The number of vertices in the largest connected component of this random network varied from 744 to 611 and 472 after 15\% and 25\% of the most central vertices were removed. Furthermore, the degree distribution of this network varied only slightly. This same experiment was repeated for the protein interaction network of \textit{Saccharomyces cerevisiae} and, after removing 15\% of the most central vertices, the size of the largest connected component reduced from 1647 to 17 vertices, hence making PINs less robust to attack than random networks.

\section{IMPORTANCE AND FUTURE DIRECTIONS}

Interactome networks have been of interest in various disciplines because they carry substantial information about the development of human disease \cite{Vidal}. Predicting protein functionality as well as choosing the best targets for drug therapy are some common applications of the topological characterization of PINs. For instance, in an up-to-date publication (2015), Azevedo et al. \cite{Azevedo} built and analyzed protein interaction networks related to chemoresistance to temozolomide (TMZ), a commonly used alkylating agent for brain cancer treatment. Their \textit{in silico} experiments of topological proved to be an efficient framework for targeting chemoresistance in cancer therapy.

\section{CONCLUDING REMARKS}

We have presented an overview of the topological characterization of complex protein-protein interaction networks. Studying the degree distribution and its influence in percolation processes is crucial for understanding and predicting functional modules in many real-world networks.

Additionally, we provided critical comparisons between the robustness of random and scale-free networks based upon state-of-the-art algorithms from network theory that have spanned a wide array of emerging subfields in biology and medicine.

In the last years, computational techniques for analyzing the structure of networked systems have become relevant in many branches of biology and have impulsed the creation of new tools that will hopefully serve as a starting point for future developments in the biomedical sciences.

\section*{\textit{Acknowledgements}}

We wish to thank all people at Instituto Nacional de Bioingeniería for great discussions and the organizing committee members for their support at all stages of the research project. We would also like to thank reviewers for helpful comments regarding the preparation of this manuscript.

\end{document}